\documentclass[aps,amssymb,notitlepage]{revtex4-1}
\date{\today}
\usepackage{graphicx}
\usepackage{color}

\usepackage{array}
\usepackage{graphicx}
\usepackage{amssymb}
\usepackage{longtable}
\usepackage{amsmath,amssymb}
\usepackage{natbib}
\usepackage{bm}
\begin{document}
\bibliographystyle{h-physrev}

\title{A short explanation of the Duflo-Zuker mass model}
\author{Chong Qi}
\thanks{Email: chongq@kth.se}
\affiliation{Department of Physics, Royal Institute of Technology (KTH), Alba Nova University Center,
SE-10691 Stockholm, Sweden}\date{\today}

\begin{abstract}
In this note we give a short introduction to the Duflo-Zuker shell model mass formula which has been shown to have great prediction power but is mostly used as a black box due to its complex nature.
\end{abstract}

\maketitle
\section{Introduction}
The microscopically inspired Duflo and Zuker (DZ)  shell model mass formula \cite{dz,Duflo1995} is constructed starting from a shell-model monopole Hamiltonian and has been shown to have good extrapolation properties. It is based on the sequential filling of a pre-assumed shell structure. 
The binding energy can 
be written as,
\begin{equation}
BE=<H_m> - E_C-E_{sym} + E_P,
\end{equation}
where the monopole Hamiltonian represents an averaged
mean field extracted from the interacting shell model. $E_C$, $E_{sym}$ and $E_P$ are the microscopic Coulomb, symmetry and pairing energies, respectively. 

In this short note we would like to give a short description on each terms in the DZ mass formulae. There are two main families of the DZ model: The DZ10 formula with 10 terms as well as the full DZ33 model with 33 terms and its variety of simplified versions. We will follow closely the codes which can be found in Ref. \cite{dz} and will not touch the physics hidden behind. 

\subsection{The shell structure}
The total number of neutrons (or protons) are given as
\begin{equation}
N=\sum_{p=0}n_p,
\end{equation}
where $p$ is the principal quantum number of the (proton or neutron) Harmonic Oscillator major shell, $n_p$ is the total number of particles in the major shell. $D_p=(p+1)(p+2)$ is the degeneracy. We have $n_p=D_p$ for a fully-filled shell. 

Each major shell $p$ is separated into two parts, the orbital with the largest $j$, with $j=p+1/2$ and degeneracy $D_j=2(p+1)$, and the rest part with the degeneracy $D_r=p(p+1)$ \cite{Duflo1995}.

\subsection{The master monopole terms}
The monopole Hamiltonian is defined as
\begin{equation}\label{dz}
H_m= H_M+H_s + H_d,
\end{equation}
where $H_M$ is a macroscopic term involving all nucleons. 
The microscopic spherical $H_s$ and deformed $H_d$ parts of the Hamiltonian take into account the residual correlation of 
valence nucleons in the open shell. 

The main master terms of $H_M$ are given as
\begin{eqnarray}
M+T&=& \frac{1}{2\rho} \left[\left(\sum_{p=0}
  \frac {m_p}{\sqrt{D_p}}\right)^2+ \left(\sum_{p=0}
  \frac {t_p}{\sqrt{D_p}}\right)^2\right],
\end{eqnarray}
$\rho=A^{1/3}[1-\left(\frac{T}{A}\right)^2]^2$ is a scaling factor.
As noted in Ref. \cite{Duflo1995}, the fitting to experimental binding energies favors $M+T$ rather than $M-T$. Above equation can be rewritten as the summation over the contributions from the protons and the neutrons as
\begin{eqnarray}\label{eq5}
M+T&=&\frac{1}{\rho} \left[\left(\sum_p
  \frac {n_p}{\sqrt{D_p}}\right)^2+ \left(\sum_p
  \frac {z_p}{\sqrt{D_p}}\right)^2\right],
\end{eqnarray}
where $n_p$ and $z_p$ denote the number of neutrons and protons in the major shell. We have $m_p=n_p+z_p$ and $t_p=|n_p-z_p|$.
The master Hamiltonian $M+T$ and its surface term $(M+T)/\rho$ resemble the volume and surface energies of the liquid drop model but contain a strong HO shell effect. This is compensated by other terms.

\subsection{Deformation}
In the spherical case of the DZ mass model, the binding energy is calculated by assuming normal filling of the proton and neutron orbitals. Deformation is simply defined as the promotion of four protons and four neutrons to the next major shell. For deformed nuclei, the quadrupole correlation energy thus gained through $H_d$ may eventually offset the loss of spherical monopole energy. In practice,  for nuclei with $Z>50$, both sphere and deformation calculations are done and the lower binding energy is kept.

\subsection{The symmetry energy}
The nuclear symmetry
energy describes the response of the nucleus to the neutron-proton number asymmetry and is closely related to the isovector channel of the nuclear interaction.
In the DZ model the symmetry energy is defined by a two-term expression as
\begin{equation}
E_{sym}=a_{sym}\frac{T(T+1)}{A\rho}-a_{ssym}\frac{T(T+1)}{A^{4/3}\rho^2},
\end{equation}
where the first and second terms are the symmetry energy and surface symmetry energy, respectively.
It should be mentioned that the symmetry energy in the DZ model is defined in a way that is different from that of the standard liquid drop model which is proportional to $(N-Z)^2$ instead. Moreover, other monopole terms of the DZ model also contribute to the symmetry energy. In particular, the two master terms also contribute to the symmetry energy in a significant way, which mostly take positive values. This can be understood by looking at the definition of the master term $M+T$ in Eq. (\ref{eq5}) which is proportional to the sum of the squares of proton and neutron numbers.
This is contrary to the liquid drop model where the volume energy term only depend on the mass number.

\subsection{The Coulomb energy}
The Coulomb energy is given as
\begin{equation}
E_C=a_C\frac{-Z(Z-1)+0.76[Z(Z-1)]^{2/3}}{A^{1/3}[1-\left(\frac{T}{A}\right)^2]},
\end{equation}
which scales in a way that is slightly different from other terms.

\section{The simplified ten-term DZ mass model  (DZ10)} 
There are several different versions of the DZ model available \cite{Duflo1995}.
The original DZ10 model contains four macroscopic terms, including the Coulomb energy, symmetry and surface symmetry energies and the pairing energy, and six monopole terms. Studies on the different terms of the DZ10 model can also be found in Ref. \cite{Mendoza-Temis2010,Bar2012a}.

Besides the master term $M+T$ (or FM+ as denoted in Ref. \cite{Duflo1995}) and its surface part, the DZ10 model contains three spherical monopole terms, one deformed monopole terms and one spin-orbital $S$ term which is added as a correction to the master term. It means that one need nine parameters in total for the discription of 'spherical' nuclei and seven parameters for the description of 'deformed' nuclei.

The spherical term $H_s$ in Eq.~(\ref{dz}) is given as
\begin{equation}\label{dz_sph}
<H_{s}>= \frac{1}{\rho} \left[a_{s} {S_{3}} +
b_s \frac{S_{3}}{\rho} +c_sS_{4}\right],
\end{equation}
where  $a_s$, $b_s$ and $c_s$ are constants to be determined. 
The operators are given as
\begin{equation}
S_3=\frac{n_{\nu}\bar{n}_{\nu}(n_{\nu}-\bar{n}_{\nu})}{N_{\nu}}+
\frac{n_{\pi}\bar{n}_{\pi}(n_{\pi}-\bar{n}_{\pi})}{N_{\pi}},
\end{equation} 
and
\begin{equation}
S_4=
  \frac{2^{\sqrt{p_{\nu}}}n_{\nu}\bar{n}_{\nu}}{N_{\nu}}\cdot
  \frac{2^{\sqrt{p_{\pi}}}n_{\pi}\bar{n}_{\pi}}{N_{\pi}}, 
\end{equation}
where $n$ ($\bar{n}=N-n$) defines the number of valence nucleons (holes) in the open spin-orbital major shell and $N$ is the corresponding degeneracy. The expectation value of the Hamiltonian $H_s$ is calculated by assuming the normal filling scheme of nucleons.

It should be mentioned that the SO major shell is composed of the largest $j$ orbital of the HO major shell $p$ and the remaining part of the lower major shell $p-1$. For examples, the first, second and third SO major shells are $0s_{1/2}$, $0p_{3/2}$ and $0p_{1/2}-0d_{5/2}$, respectively. Concerning the degeneracy, we have $N_p=D_j(p)+D_r(p-1)=p(p+1)+2$. A schematic picture of the HO and SO major shells may be found in the original paper.

The deformed Hamiltonian $H_d$ can be constructed in a way  similar to $S_4$ but takes into account the effect of the promotion of valence nucleons to
the next shell~\cite{Duflo1995},
\begin{equation}
<H_{d}>=\frac{d}{\rho}\left[\left(\frac{n'_{\nu}\bar{n'}_{\nu}}{N^{3/2}_{\nu}}\right) \cdot
\left(\frac{n'_{\pi}\bar{n'}_{\pi}}{N^{3/2}_{\pi}}\right)\right],
\end{equation}
where $n'=n-u$, $u=4$ is the number of particles that are promoted to the next major shell and $d$ is the constant.

The DZ10 mass model contains a rather sophisticated $S$ (or SO) term which is added to the master term $M$. The competition between $M$ and $S$ is responsible for changing the shell structure from HO to spin-orbital ones with $N (Z) = 28, 50, 82, 126$ and 184 \cite{Zuker2008}. One has
\begin{eqnarray}
S_{\nu}=\sum_{p=0}(n_{jp}p/2-n_{rp})\nonumber\\
\times\left[(1+\frac {n_p}{\sqrt{D_p}})\frac{p^2}{D_p^{3/2}}+(1-\frac {n_p}{\sqrt{D_p}})\frac{4p-5}{D_p^{3/2}}\right].
\end{eqnarray}

In the original DZ10 model, the pairing energy is expressed as
\begin{equation}
E_P=a_p\frac{\varepsilon_p}{\rho},
\end{equation}
where 
\begin{equation}
\varepsilon_p=\left\{\begin{array}{l}
|N-Z|/\rho, ~~~\rm{for~odd-odd ~nuclei}\\
2-|N-Z|/\rho, ~~~\rm{for~even-even ~nuclei}\\
1-|N-Z|/\rho, ~~~\rm{for~N>Z~even-N-odd-Z ~nuclei}\\
1, ~~~\rm{for~N>Z~even-Z-odd-N ~nuclei}\\
1-|N-Z|/\rho, ~~~\rm{for~N<Z~even-Z-odd-N ~nuclei}\\
1, ~~~\rm{for~N<Z~even-N-odd-P ~nuclei}
\end{array}\right.
\end{equation}

The term $2T/A\rho$ and the correction term to the surface symmetry energy and the pairing energy of the forms $T(T-1/2)/A\rho^4$ (denoted as 'Wigner energy' in Ref. \cite{dz}) have negligible influence on the global description of the available binding energies (see, e.g., Ref. \cite{Qi2012436}). 

A simplified version may be given as
\begin{equation}\label{pair}
E_P=a_p\frac{2-v}{\rho},
\end{equation}
where $v$ denotes the seniority of the nucleus. In the present work the seniority quantum number is assumed to be zero for the ground states of even-even nuclei, one for those of odd-$A$ nuclei and two for odd-odd nuclei with isospin $T=|N-Z|/2$ \cite{Qi2012436}. The seniority is assumed to be zero for the $T=1$ ground states of odd-odd $N=Z$ nuclei.

\subsection{The full DZ mass model} 
The full DZ33 mass model contains 28 monopole terms, which include
FM+, fm+,
FS+,  fs+, FS-, fs-,
FC+, fc+, PM+,  pm+, 
PS+, ps+, PS-, ps-, 
S3,  s3,  SQ-, sq-,
D3,  d3,  QQ+,  qq+,  
D0,  d0, QQ-, qq-,  
SS,  ss,
as well as the Coulomb energy, symmetry energy, surface symmetry energy and two pairing terms \cite{Duflo1995} .
The volume terms (capital letters) and the surface terms (small letters). S3,  s3,  SQ- and sq- are the valence spherical terms while D3,  d3,  QQ+,  qq+,  
D0,  d0, QQ- and qq- are deformed terms. A short explanation of the different terms is also given in Ref. \cite{bertsch}. Here we give the expressions for above terms following the original code as (Please be aware that the extra scaling $1/\rho$ is not given explicitly for simplicity) 
\begin{equation}
\begin{array}{lll}
FS+&=&(\sum_{p}\frac{s_{\pi}+s_{\nu}}{p+1})^2+(\sum_{p}\frac{s_{\pi}-s_{\nu}}{p+1})^2\\
&=&2(\sum_{p}\frac{s_{\pi}}{p+1})^2+2(\sum_{p}\frac{s_{\nu}}{p+1})^2\\
FS-&=&(\sum_{p}\frac{s_{\pi}+s_{\nu}}{p+1})^2-(\sum_{p}\frac{s_{\pi}-s_{\nu}}{p+1})^2\\
&=&4\left(\sum_{p}\frac{s_{\pi}}{p+1}\right)\left(\sum_{p}\frac{s_{\nu}}{p+1}\right)\\
FC+&=&\left(\sum_p
  \frac {m_p}{D_p}\right)\left(\sum_{p}\frac{s_{\pi}+s_{\nu}}{(p+1)D_p^{1/2}}\right)\\
  &&+\left(\sum_p
  \frac {t_p}{D_p}\right)\left(\sum_{p}\frac{s_{\pi}-s_{\nu}}{(p+1)D_p^{1/2}}\right)\\
PM+&=&\left(\sum_p
  \frac {n_{\pi p}}{D_p^{1/4}}\right)^2+ \left(\sum_p
  \frac {n_{\nu p}}{D_p^{1/4}}\right)^2\\
PS+&=&2(\sum_{p}\frac{s_{\pi}D_p^{1/4}}{(p+1)})^2+2(\sum_{p}\frac{s_{\nu}D_p^{1/4}}{(p+1)})^2\\
 PS-&=&4\left(\sum_{p}\frac{s_{\pi}D_p^{1/4}}{(p+1)}\right)\left(\sum_{p}\frac{s_{\nu}D_p^{1/4}}{(p+1)}\right)\\
 \end{array}
\end{equation}
where $s=n_{jp}p/2-n_{rp}$
and
\begin{equation}
\begin{array}{lll}
S3,&&\rm{Same ~as~ DZ10}\\
 SQ-&=&\frac{2n_{\nu}\bar{n}_{\nu}}{N_{\nu}}
  \frac{2n_{\pi}\bar{n}_{\pi}}{N_{\pi}}\\
D3&=&\frac{n'_{\nu}\bar{n'}_{\nu}(n'_{\nu}-\bar{n'}_{\nu})}{N_{\nu}^2}+
\frac{n'_{\pi}\bar{n'}_{\pi}(n'_{\pi}-\bar{n'}_{\pi})}{N_{\pi}^2}\\
QQ+&=&\frac{2(n'_{\nu}\bar{n'}_{\nu})^2}{N^3_{\nu}}+
  \frac{2(n'_{\pi}\bar{n'}_{\pi})^2}{N^3_{\pi}}\\\\
D0&=&16-QQ-\\
QQ-&=&\frac{2n'_{\nu}\bar{n'}_{\nu}}{N^{3/2}_{\nu}}
  \frac{2n'_{\pi}\bar{n'}_{\pi}}{N^{3/2}_{\pi}}.\\
\end{array}
\end{equation}

There is an extram contribution from $SS$ term to nuclei with proton and/or neutron numbers larger than 20. The term $SS$ is given as
\begin{equation}
\begin{array}{lll}
SS&=&\sum_{p>2}ss_{\pi,p}\left(  \frac {n_{\pi p}}{D_p^{1/2}} +\frac{s_{\pi}}{p+1}  \right) + \sum_{p>2}ss_{\nu,p}\left(  \frac {n_{\nu p}}{D_p^{1/2}} +\frac{s_{\nu}}{p+1}  \right),
\end{array}
\end{equation}
where we have
\begin{equation}
ss_p=\left\{\begin{array}{lll}
n_j(p-1)/2p ~~~\rm{if~~~} n_j\le p(p-1)\\ 
(p-1)^2/2-(n_j-p(p-1))/p ~~~\rm{if~~~} n_j> p(p-1)
\end{array}\right.
\end{equation}

Terms like FM-, FC- and PM- and PC$\pm$ as well as their surface terms are not considered. 

It should be mentioned that DZ10 is not a simplified version of the DZ33 model. They contain different monopole terms. In particular, the $S$ term is only present in the DZ10 model.
The shell structure is assumed to be the same in both models.

The pairing energy is slightly different than that in DZ10 in the sense that it is given as a negative contribution relative to the even-even ones as
\begin{equation}
\varepsilon_p=\left\{\begin{array}{l}
-2, ~~~\rm{for~odd-odd ~nuclei}\\
-2|N-Z|/\rho, ~~~\rm{for~even-even ~nuclei}\\
-1-2|N-Z|/\rho, ~~~\rm{for~N>Z~even-N-odd-Z ~nuclei}\\
-1-|N-Z|/\rho, ~~~\rm{for~N>Z~even-Z-odd-N ~nuclei}\\
-1-2|N-Z|/\rho, ~~~\rm{for~N<Z~even-Z-odd-N ~nuclei}\\
-1-|N-Z|/\rho, ~~~\rm{for~N<Z~even-N-odd-P ~nuclei}
\end{array}\right.
\end{equation}
In the other words, the pairing energy in DZ33 is shifted by $2+|N-Z|/\rho$ relative to that in DZ10.

The full DZ33 model contains an extra term for nuclei with the valence protons and/or neutrons in the remaining $r$ orbitals of the HO major shell.
\begin{equation}\label{pair}
E_P'=a_p'(\frac{\delta_{\nu}}{A}+\frac{\delta_{\pi}}{A}),
\end{equation}
where $\delta=1$ if the valence nucleons occupy $r$ and 0 if they occupy $j$. It is denoted as a pairing term in the original code but it does not show any odd-even staggering.

\section{DZ model with reduced terms}
As indicated in above discussions, one may have even more combination fromt the monopole terms. Many of them are disfavored in the fitting to the available data. In the original paper, only 28 terms are considered. Even simplier versions may be proposed without significant deterioration of is performance.
In Ref. \cite{Qi14}, only the following 15 monopole terms are considered:
FM+ ,   
fm+  ,  
FS  ,   
fs- ,   
fc+ ,   
PM+ ,   
PS+ ,   
S3 ,    
s3 ,    
SQ- ,   
sq- ,   
d3+QQ+ ,
D0 ,    
d0 and    
SS.


\end{document}